\documentstyle[prd,aps,epsf,floats,amsfonts,amssymb,amsmath]{revtex}

\begin{document}
\title{Lorentz Invariance Breakdown and Constraints from Big-Bang Nucleosynthesis}
\author{G. Lambiase$^{a,b}$}
 \address{$^a$Dipartimento di Fisica "E.R. Caianiello" Universit\'a di Salerno, 84081 Baronissi (Sa),
 Italy,}
 \address{
 $^b$INFN - Gruppo Collegato di Salerno, Italy.}
\maketitle
\begin{abstract}
The Standard Model Extension formulated by Colladay and
Kosteleck\'y is reviewed in the framework of the $^4He$ primordial
abundance. Upper bounds on coefficients for the Lorentz violation
are derived using the present observational data.
\end{abstract}


 \maketitle

Many attempts aimed to construct a quantum theory of gravity have
shown that spacetime might have a non trivial topology at the
Planck scale. However, the difficulties to build up a complete
theory of quantum gravity have motivated the development of
semiclassical approaches in which CPT and Lorentz invariance
breakdown occur at the level of effective theory. As suggested by
Kosteleck\'y and Samuel \cite{kostelecky,k-potting}, a departure
from Lorentz invariance could manifest itself as an effect of
non-locality in string theory: The interactions among tensor
fields give rise to non-zero expectation values for the Lorentz
tensors that induce spontaneously broken Lorentz symmetry. The
general effective field theory describing these Lorentz and CPT
violations is the Standard Model Extension (SME) \cite{calladay}.
This model, along with the usual SM and gravitational Lagrangian,
includes all possible coordinate-invariant terms constructed with
SM, gravitational fields, and violating Lorentz symmetry
\cite{kostelecky-gravity}.

The aim of this paper is to derive, in the framework of Big-Bang
Nucleosynthesis (BBN), bounds on parameters of SME. BBN is a
cornerstone of standard cosmology: with cosmic background
radiation, it provides a strong evidence that during the early
phases - i.e. between a fraction of seconds ($\sim 0.01$ s) and
few hundred seconds after the BB - the Universe was hot and dense.
BBN describes the sequence of nuclear reactions leading to the
synthesis of light elements. Our analysis follows the paper by
Bernstein, Brown and Feinberg \cite{bernstein}, in which the model
of helium synthesis in the early Universe is discussed. During the
BBN phase, the geometry of the expanding early Universe is
described by the Friedman-Robertson-Walker (FRW) metric
$ds^2=dt^2-a^2(t)(dX^2+dY^2+dZ^2)$. $a(t)$ is the scale factor
(the spatial curvature has been taken equal to zero). The
dynamical equation for the evolution of the scale factor is
$H^2=\frac{8\pi G}{3}\rho$, where $H={\dot a}/a$ is the Hubble
parameter. The corrections to General Relativity discussed by
Kosteleck\'y and Bluhm \cite{kostelecky-gravity} are not
considered here.

As shown by Kosteleck\'y and Lehnert \cite{lehnert},
the dispersion relation of relativistic fermions is given by
\begin{widetext}
\begin{equation}\label{relaz-disp2}
  p^\mu p_\mu-m^2=-2c^{\mu\nu}p_\mu p_\nu +2a^\mu p_\mu +2 a_\mu c^{\mu\nu}p_\nu
  -c_{\mu\nu}c^{\mu\sigma}p^\nu p_\sigma -a^2-2m e_\mu p^\mu
  +e_\mu e_\nu p^\mu p^\nu +P^2+A^2
\end{equation}
 \[
\pm 2 \sqrt{(V_\mu A^\mu)^2-V^2
  A^2+(PT^{\mu\nu}-S{\tilde
  T}^{\mu\nu})(-2V_\mu
  A_\nu+T_{\mu\alpha}T^\alpha_{\phantom{\alpha}\nu})+
  10(T^{\alpha\beta}T_{\alpha\beta})^2+9(T^{\mu}_{\phantom{\mu}\alpha}T^{\alpha\nu})^2
  }\,.
 \]
\end{widetext}
where $\pm$ signs take care of the fermion helicities, and
\begin{eqnarray}
  V^\mu &=& p^\mu+c^{\mu\nu}p_\nu-a^\mu\,, \quad
  A^\mu=d^{\mu\nu}p_\nu -b^\mu\,, \quad
  S=e^\mu p_\mu-m\,, \quad P=f^\mu p_\mu \,, \nonumber \\
  T^{\mu\nu}&=&\frac{1}{2}\, g^{\mu\nu}_{\phantom{\mu\nu}\rho}p^\rho-\frac{H^{\mu\nu}}{2}\,,
  \quad
  {\tilde T}^{\mu\nu}=\frac{1}{2}\,
  \varepsilon^{\mu\nu\alpha\beta}T_{\alpha\beta}\,. \nonumber
\end{eqnarray}
%
All coefficients violate particle Lorentz invariance, and may be
flavor depending. Besides, $a_\mu$, $b_\mu$, $e_\mu$, $f_\mu$,
$g_{\lambda \mu\nu}$ break CPT. As in \cite{KPLB}, we assume that
the only non-vanishing coefficients are $c_{\mu\nu}$ (where
$c_{\mu\nu}\eta^{\mu\nu}=0$). Due to the symmetries of the FRW
metric, we also assume that: {\bf a)} The spatial coefficients
$c_{ij}$ are isotropic, $c_{ij}\simeq c\,\delta_{ij}$, where
$c\simeq c_{XX}\simeq c_{YY}\simeq c_{ZZ}$ ($c=c_{00}/3$). {\bf
b)} The off-diagonal elements of the coefficients $c_{\mu\nu}$
vanish, i.e. $c_{0i}=0$,\,\, $i=X, Y, Z$. We therefore get
$p\simeq {\cal A}\, \sqrt{E^2-{\tilde m}^2}$, where ${\cal
A}\equiv \left(1+\frac{8c_{00}}{3}\right)^{1/2}$, and $ {\tilde
m}=m\left(1-c_{00}\right)$. Moreover, we choose as conventional
the photon sector \cite{photonconv}, thus $E_\gamma=k$.

The energy density of relativistic particles ($T\gg m, \mu$, where
$\mu$ is the chemical potential) filling up the early Universe is
given by $ \rho=\frac{g_s}{(2\pi)^3}\int E\, n_{E}\, d^3 p$
\cite{statistics}, where $g_s$ represent the degeneracy factors
for particle species involved ($g_\gamma=2$, $g_e=4$, $g_\nu=2$),
and $n_E$ is the number density of particles. Since $c^{e,
\nu}_{00}\ll 1$, the total energy density can be written as
$\rho=\rho_b+\rho_f\simeq \rho^{(0)}+\rho^{(LIV)}$, where
$\rho^{(0)}=\frac{\pi^2}{30}\,g T^4$ is the standard energy
density, and
\begin{equation}
  \rho^{(LIV)} = \frac{\pi^2}{30}\left[
  \frac{7}{2}\, c_{00}^e\, g_e+\frac{21}{2}\, c_{00}^{\nu}\, g_\nu\right]\, T^4\,. \label{rhoLIV}
\end{equation}
$g\equiv g_b+\frac{7}{8}\, g_F=\frac{43}{4}$ $(g_F=g_e+3g_\nu=10)$
is the effective number of degrees of freedom (it is implicitly
assumed that muon and tau neutrinos have a small mass compared to
the effective temperature, and that no other massless species are
present).

The formation of the primordial ${}^4 He$ occurs at the early time
when the temperature of the Universe was $T\sim 100$ MeV and the
energy and number density were dominated by relativistic
particles: leptons (electrons, positrons, neutrinos) and photons.
At this stage of the Universe evolution, the smattering of
neutrons and protons does not contribute in a relevant way to the
total energy density. All these particles are in thermal
equilibrium owing to their rapid collisions. Besides, protons and
neutrons are kept in thermal equilibrium by their interactions
with leptons
\begin{eqnarray}
 \nu_e+n &\,\, \longleftrightarrow \,\, & p+e^- \label{int1} \\
 e^++n &\,\, \longleftrightarrow \,\, & p+{\bar \nu}_e \label{int2} \\
 n &\,\, \longleftrightarrow \,\, & p+e^- +{\bar \nu}_e \label{int3}
\end{eqnarray}
To estimate the neutron abundance in the expanding Universe, one
has to compute the conversion rate of protons into neutrons,
$\lambda_{pn}(T)$, and its inverse $\lambda_{np}(T)$. At enough
high temperature, the weak interaction rate is given by
\begin{equation}\label{Lambda}
\Lambda(T)=\lambda_{np}(T)+\lambda_{pn}(T)\,.
\end{equation}
$\lambda_{np}$ and $\lambda_{pn}$ are related by $\lambda_{np}(T)=
e^{-Q/T}\lambda_{pn}(T)$, with $Q={\tilde m}_n-{\tilde m}_p$.
$\lambda_{np}(T)$ is expressed as the sum of the rates associated
to the individual processes (\ref{int1})-(\ref{int3})
\begin{equation}\label{rate}
  \lambda_{np}=\lambda_{n+\nu_e \rightarrow  p+e^-}+
  \lambda_{n+e^+ \rightarrow  p+{\bar \nu}_e}+
  \lambda_{n \rightarrow  p+e^- +{\bar \nu}_e}\,.
\end{equation}
During the freeze-out period, the following approximations can be
done \cite{bernstein}: {\bf 1.} The temperatures involved are the
same, i.e. $T_\nu=T_e=T_\gamma=T$. ${\bf 2.}$ The temperature $T$
is low with respect to the typical energies $E$ that contribute to
the integrals entering in the definition of the rates. One then
replaces the Fermi-Dirac distribution with the Boltzmann
distribution, $n_E \simeq e^{-E/T}$. ${\bf 3.}$ The electron mass
$m_e$ can be neglected as compared with the electron and neutrino
energies, i.e. $m_e\ll E_e, E_\nu$.

The processes (\ref{int1})-(\ref{int3}) are described by weak
interactions which involve the gauge boson $W$ as mediator.
According to the SME's prescription \cite{calladay}, the
interaction $npW$ may be described in terms of an effective weak
gauge theory
in which $p$ and $n$ are arranged in the doublet $\left(\begin{array}{c} n \\
p\end{array} \right)$. For the $n$-$p$ sector, one has a
coefficient for the Lorentz invariance corresponding to the Left-doublet $\left(\begin{array}{c}n \\
p\end{array} \right)_L$, and one coefficient corresponding to each
Right-component, $n_R$ and $p_R$. The combination of these
coefficients leads to the coefficients $c_{\mu\nu}^n$ for $n$,
$c^p_{\mu\nu}$ for $p$, and $c^{np}_{\mu\nu}$ for the vertex $npW$
\cite{coeff}. $c^n_{\mu\nu}$ and $c^p_{\mu\nu}$ enter also in the
propagators of $n$ and $p$, respectively. For the lepton sector
$e$-$\nu_e$, $c_{\mu\nu}^e$ and $c_{\mu\nu}^{\nu_e}$ enter in the
propagators of $e$ and $\nu_e$, respectively, whereas
$c_{\mu\nu}^e$ enter in the vertex $e\nu_eW$.

The interaction rate for the process (\ref{int1}) is
\cite{comment}

\begin{widetext}

\begin{equation}\label{ratedef}
  d\lambda_{n+\nu_e\to p+e^-}= \frac{\langle |{\cal M}|^2\rangle}{2{\tilde m}_n}
  \frac{d^3p_e}{(2\pi)^3 2E_e}\frac{d^3p_{\nu_e}}{(2\pi)^3 2E_{\nu_e}}
  \frac{d^3p_p}{(2\pi)^3 2E_p}\, (2\pi)^4
  \delta^{(4)}(p_n+p_{\nu_e}-p_p-p_e)\, n_{E_\nu}(1-n_{E_e})\,,
\end{equation}

\end{widetext}
where
\begin{equation}\label{M}
{\cal M}=\left(\frac{g_w}{8M_W}\right)^2[{\bar u}_p\Omega^\mu
u_n][{\bar u}_e\Sigma_\mu v_{\nu_e}]\,,
\end{equation}
$\Omega^\mu=(\gamma^\mu+c^{\mu\nu}_{np}\gamma_\nu)(c_V-c_A\gamma^5)$,
$\Sigma^\mu=(\gamma^\mu+c_{e}^{\mu\nu}\gamma_\nu)(1-\gamma^5)$. In
(\ref{M}), we used the fact that the transferred momentum
$q^\mu=p^\mu_n-p^\mu_p$ satisfies the condition $q^2\ll M_W^2$,
thus the boson propagator is $i\eta_{\mu\nu}M_W^{-2}$ (corrections
induced by the Lorentz violating (anti-symmetric) coefficients
$(k^A_{\phi\phi})^{\mu\nu}$ \cite{higgs} are discarded).

In the nucleosynthesis phase, the energy recoil of nucleons can be
neglected \cite{bernstein}, i.e. $p_n^\mu=(m_n, {\bf 0}),
p_p^\mu=(m_p,{\bf 0})$. After lengthy computations, Eq.
(\ref{ratedef}) becomes
%
 \[
 \lambda_{n+\nu_e\to p+e^-}={\tilde A} {\cal A}_e^3{\cal A}_{\nu_e}^3\,
  T^5\, I_y\,, \quad
   I_y = \int_y^\infty \epsilon\, (\epsilon-Q')^2\sqrt{\epsilon^2-y^2}\,  n_{\epsilon-Q'}(1-n_\epsilon)
      \, d\epsilon\,, \]
      \[
   y=\frac{{\tilde m}_e}{T}\,, \,\, Q'=\frac{{\tilde m}_n-{\tilde m}_p}{T}\,, \,\,
{\tilde A}=A(1+B)\,, \quad
  A=\frac{g_V^2+3g_A^2}{2\pi^3}=1.02 \times
   10^{-47}\mbox{eV}^{-4}\,, \]
   \[
  B=c_{00}^e+c_{00}^{\nu_e}+\frac{2(g_V^2+g_A^2)}{g_V^2+3g_A^2}\,
  (c_{00}^n+c_{00}^p)+
   \frac{2(g_V^2-g_A^2)}{g_V^2+3g_A^2}\,
  (c_{00}^{e}+c_{00}^{np})\,. \]
In a similar way, one derives
 \begin{equation}\label{rate2}
   \lambda_{n+e^+ \rightarrow  p+{\bar \nu}_e} = {\tilde A}\, {\cal A}_e^3{\cal A}_{\nu_e}^3\,
   T^5\, J_y\,, \quad
      J_y =\int_y^\infty
  \epsilon\, (\epsilon+Q')^2\,
 \sqrt{\epsilon^2-y^2}\,n_{\epsilon}(1-n_{\epsilon+Q'})\, d\epsilon \,.
 \end{equation}
Finally, for the neutron decay $n\to p+e^-+{\bar \nu}_e$ one
obtains $\tau = \lambda_{n \rightarrow p+e^- +{\bar
\nu}_e}^{-1}=(887+{\cal O}(c^{e, \nu_e, p, n}_{00}))$ sec. In
computing $\Lambda$ we shall neglect $\lambda_{n \rightarrow p+e^-
+{\bar \nu}_e}$ since $n$ may be considered stable during the BBN
phase \cite{bernstein}. The approximations ${\bf 1.} - {\bf 3.}$
imply $\lambda_{n+e^+\to p+{\bar \nu}_e}=\lambda_{n+\nu_e\to
p+e^-}$.

The rate (\ref{rate2}) can be written as
\begin{equation}\label{l=l0+lLIV}
  \lambda_{n+e^+\to p+{\bar \nu}_e}=\lambda_{n+e^+\to p+{\bar \nu}_e}^{(0)}+
  \lambda_{n+e^+\to p+{\bar \nu}_e}^{(LIV)}\,,
\end{equation}
where $\lambda_{n+e^+\to p+{\bar \nu}_e}^{(0)} = AT^3(4!
T^2+2\times 3! QT+2!Q^2)$ is the standard result \cite{bernstein},
and
\[
      \lambda_{n+e^+\to p+{\bar \nu}_e}^{(LIV)} = 4A\,T^3Q^2\left[4!\Big(c_{00}^e+c_{00}^{\nu_e}+\frac{B}{4}\Big)
      \frac{T^2}{Q^2}+2\,\, 3!\Big(c_{00}^e+c_{00}^{\nu_e}+\frac{B}{4}+\frac{Q_m}{4Q}\Big)\frac{T}{Q}+
     2!\Big(c_{00}^e+c_{00}^{\nu_e}+\frac{B}{4}+\frac{Q_m}{2Q}\Big)
     \right]\,.
\]
Here $Q_m=m_pc_{00}^p-m_nc_{00}^n$. From (\ref{Lambda}) and
(\ref{rate}), it follows $\Lambda(T) \simeq
2\lambda_{np}=\Lambda^{(0)}+\Lambda^{(LIV)}$ ($\Lambda^{(\alpha)}
= 4\,\lambda_{n+e^+\to p+{\bar \nu}_e}^{(\alpha)}$).

To estimate the primordial mass fraction of ${}^4He$ we employ the
expression \cite{kolb}
\begin{equation}\label{Yp}
  Y_p=\lambda\left(\frac{2x(t_f)}{1+x(t_f)}\right)\,,
\end{equation}
where $\lambda = e^{-(t_n-t_f)/\tau}$, $t_f$ and $t_n$ are the
time of freeze-out of the weak interactions and of the
nucleosynthesis, respectively, and $x(t_f)=\exp[-Q/T(t_f)]$ is the
neutron to proton equilibrium ratio. $\lambda(t_f)$ is the
fraction of neutrons that in the time $t\in [t_f, t_n]$ decays
into protons. At the radiation era \cite{kolb,bernstein},
$T(t)\sim \left(t/\mbox{sec}\right)^{-1/2}\mbox{MeV}$, one obtains
the deviation from the fractional mass $Y_p$ due to the variation
of the freezing temperature $T_f$
 \begin{equation}\label{deltaYp}
\delta Y_p=
Y_p\left[\left(1-\frac{Y_p}{2\lambda}\right)\ln\left(\frac{2\lambda}{Y_p}-1\right)-\frac{2t_f}{\tau}\right]
\frac{\delta T_f}{T_f}\,.
 \end{equation}
In (\ref{deltaYp}) we set $\delta T_n = \delta T(t_n)=0$ being
$T_n$ fixed by the deuterium binding energy \cite{diego}. From
$\Lambda \simeq H = \sqrt{\frac{8\pi G}{3}\, \rho}$, one derives
the freeze-out temperature $T=T_f\left(1+\frac{\delta
T_f}{T_f}\right)$, where $T_f\sim 0.6$ MeV, and
\begin{equation}\label{deltaTsplit}
  \frac{\delta T_f}{T_f} \simeq -1.83\,c_{00}^e-1.7\, c_{00}^{\nu_e}+6.5 \times
  10^{-3}c_{00}^{np}+
  236.16\, c_{00}^n-236,61 \, c_{00}^p\,.
  \end{equation}
Estimations on mass fraction $Y_p$ of baryons converted to $^4He$
during BBN are reported in Table I. By using $Y_p=0.2476$ and
$|\delta Y_p| < 10^{-4}$, Eq. (\ref{deltaYp}) gives
\begin{equation}\label{deltaYp<}
  \left|\frac{\delta T_f}{T_f}\right| < 4.7 \times 10^{-4}\,.
\end{equation}
Eqs. (\ref{deltaTsplit}) and (\ref{deltaYp<}) lead to upper bounds
on coefficients for the Lorentz violation
\[
  |c_{00}^e+0.93\, c_{00}^{\nu_e}-3.5\times 10^{-3}c_{00}^{np}+129.3\,
  c_{00}^p-129.05\, c_{00}^n| < 2.56 \times 10^{-4}\,.
\]


{\it Acknowledgments.} It is a pleasure to thank V.A. Kosteleck\'y
for discussions and valuable remarks, D.F. Torres for comments,
and A. Iorio for reading the paper. Research supported by PRIN
2004.

\begin{table}
\caption{\label{I}{Values of $Y_p$ and $\delta Y_p$.}}
\begin{tabular}{ccc}
 \multicolumn{1}{c}{$Y_p$}  \hspace{0.7in} & \multicolumn{1}{c}{$\delta Y_p$} \hspace{0.7in}
 & \multicolumn{1}{c}{{Ref. } }  \\ \hline
 0.2476 \hspace{0.7in} &  $\pm$ 0.0010 \hspace{0.7in} &  \cite{kirkman} \\
 0.2452 \hspace{0.7in} &  $\pm$ 0.0015 \hspace{0.7in} & \cite{izatov99} \\
 0.2429 \hspace{0.7in} &  $\pm$ 0.0009 \hspace{0.7in} & \cite{izotov04} \\
 0.2479 \hspace{0.7in} &  $\pm$ 0.0004 \hspace{0.7in} & \cite{coc} \\
 0.238 \hspace{0.7in} &  $\pm$ 0.002 \hspace{0.7in} &  \cite{fields} \\
 0.244 \hspace{0.7in} &  $\pm$ 0.002 \hspace{0.7in} &  \cite{izotov} \\
 0.234 \hspace{0.7in} &  $\pm$ 0.003 \hspace{0.7in} &  \cite{olive} \\
\end{tabular}
\end{table}


\begin{thebibliography}{99}
\bibitem{statistics} In computing the energy density $\rho$ we take into account results of
                    \cite{stat-term}. There it is shown that the formalism of statistical mechanics in the presence
                    of Lorentz breakdown is the same as for
                    conventional statistical mechanics. The same holds for the laws of
                    thermodynamics, although the specific expectation values of
                    thermodynamical quantities may be modified owing to corrections induced by
                    coefficients breaking Lorentz symmetry.
\bibitem{photonconv} As discussed in literature \cite{JMP}, a
    suitable redefinition of coordinates and fields allows to move the
    Lorentz invariance parameters from a sector to another sector. One
    may choose to take a sector as conventional, and shift all the
    Lorentz violation coefficients in the remaining sectors.
    We assume the photon sector as conventional.
\bibitem{coeff} In the SME, the Lagrangian density for quarks is constructed by
    adding to the usual SM the most general CPT-violating terms
    \cite{calladay}, which encode the coefficients $(c_Q)_{\mu\nu AB}$,
    $(c_U)_{\mu\nu AB}$ and $(c_D)_{\mu\nu AB}$ ($A, B=1,2,3$). The coefficients $(c_Q)_{\mu\nu AB}$ fix
    the propagator and the interactions with gauge fields of
    left-handed quarks, while $(c_U)_{\mu\nu AB}$, and $(c_D)_{\mu\nu
    AB}$ fix the propagator and the interactions of right-handed
    quarks. The combination of the coefficients $(c_Q)_{\mu\nu 11}$
    and $(c_U)_{\mu\nu 11}$ enters in the propagator of the quark $u$,
    while $(c_Q)_{\mu\nu 11}$ and $(c_D)_{\mu\nu 11}$ enter in the
    propagator of the quark $d$. The  propagators of
    bounded states of quarks (constituting neutrons and protons)
    contain combinations of these coefficients.
    In the effective weak gauge theory, we denote them with $c_{\mu\nu}^p$ and
    $c_{\mu\nu}^n$. Moreover, the vertex of the interaction
    $duW$ (hence $npW$) is controlled by the coefficients
    $(c_Q)_{\mu\nu 11}$. We set $(c_Q)_{\mu\nu 11}\equiv c_{\mu\nu}^{np}$.
\bibitem{comment} A comment is in order.
    The available phase space for the final fermion state is
    \cite{KPLB} $d\Pi = d^3p/(2\pi)^32EN(p)$. In computing the cross
    sections or decay rates, the normalization factors $N(p)$ cancels
    out with ones coming from the transition probability $\langle
    |{\cal M}|^2\rangle$, once the completeness relations of spinors
    are used. A boost of $d\Pi$ to another inertial frame is typically
    non trivial since one has to consider the redefinition of the
    spinor $\psi\to (1-c_{\mu 0}\gamma^0\gamma^\mu/2)\psi$.
    Nevertheless, in our context, we are setting $c_{0i}=0$, so that
    the spinor $\psi$ is redefined up to a constant factor, which does
    not alter the final results.
\end{thebibliography}
\end{document}